\documentclass[]{elsart}

\usepackage[thinspace,squaren]{SIunitsNEW}
\usepackage{graphicx,amssymb,amsmath}
\usepackage[square,comma,numbers,sort&compress]{natbib}
\usepackage{amsfonts}
\usepackage{psfrag}
\usepackage{graphics}
\def\ea{\epsilon_{\sf a}}
\def\eb{\epsilon_{\sf b}}
\def\nua{\nu_{\sf a}}
\def\nub{\nu_{\sf b}}
\def\Ea{E^{\rm Y}_{\sf a}}
\def\Eb{E^{\rm Y}_{\sf b}}
\def\xa{x_{\sf a}}
\def\xb{x_{\sf b}}
\def\ddd{d_{33}^{\sf e}}
\def\aaa{a_{33}^{\sf e}}
\def\full{\protect\mbox{------}}

\def\dash{{\protect\mbox{--\ --\ --}}}
\def\dotted{{\protect\mbox{${\mathinner{\cdotp\cdotp\cdotp\cdotp\cdotp\cdotp}}$}}}
\begin{document}
\begin{frontmatter}
\journal{Journal of Electrostatics}
\title{Modeling electro-mechanical properties of layered electrets: {Application of the finite-element method}}

\author{Enis Tuncer\corauthref{cor},}
\corauth[cor]{Corresponding author.}
\ead{enis.tuncer@physics.org}
\author{Michael Wegener,}
\author{Reimund Gerhard-Multhaupt} 
\address{Applied Condensed-Matter Physics, Department of Physics, University of Potsdam, D-14469 Potsdam Germany}

%\date{\today}
%\volume{}

\begin{abstract}
  We present calculations on the deformation of two- and three-layer electret systems. The electrical field is coupled with the stress-strain equations by means of {\em the Maxwell stress tensor}. In the simulations, two-phase systems are considered, and intrinsic relative dielectric permittivity and Young's modulus of the phases are altered. The numerically calculated electro-mechanical activity is  compared to an analytical expression. Simulations are performed on two- and three-layer systems. Various parameters in the model are systematically varied and their influence on the resulting piezoelectricity is estimated. In three-layer systems with bipolar charge, the piezoelectric coefficients exhibit a strong dependence on the elastic moduli of the phases. However, with mono-polar charge, there is no significant piezoelectric effect. A two-dimensional simulation illustrated that higher piezoelectricity coefficients can be obtained for non-uniform surface charges and low Poisson's ratio of phases. Irregular structures considered exhibit low piezoelectric activity compared to two-layer structures.
\end{abstract}
\end{frontmatter}

\maketitle

\section{Introduction}
\label{sec:introduction}

Charged dielectric films with soft elastic properties and at least one free (unclamped) surface can be used as electro-mechanical and electro-acoustic transducers\cite{ReimundRev,ReimundPhysTdy}. Traditional materials for such applications have been inorganic crystalline substances\cite{Cady12}. The electro-mechanical effect (piezoelectricity) observed in organic polymeric materials differs in several aspects from that of traditional inorganic  piezoelectric materials. Apart from morphological differences, piezoelectricity in polymers is based either on (i) oriented molecular dipoles (domains) or on (ii) trapped charges whose distribution breaks the symmetry inside the non-uniform anisotropic material. The underlying mechanisms of the phenomenological piezoelectricity in these materials have previously been presented in the literature\cite{Lewiner,Wintle1989,Kac0,Lekkala2000a}. Recently, several potential candidates for large piezoelectric effects in heterogeneous or porous polymer systems are reported\cite{Reimund2000b,Reimund2000a,Sessler1999,Lindner,Bauer2000b,Lekkala2000,Lekkala2001}. 
%These composite materials exhibiting quasi-piezo\-elect\-ricity are candidates as ``metamaterials'' since the piezoelectric phenomena is not observed in the constituent materials. 
Although there exist analytical models for modeling simple geometries and uniform charge distributions,  numerical solutions may be preferable for complex geometries\cite{TorquatoBook,Tuncer2002b,Tuncer2002a} and non-uniform charge distributions. In this paper, the finite-element (FE) method is first applied to solve the electric field in two- and three-layered structures with interface, surface or volume charges. We have considered non-uniform charge distributions and irregular geometries as two-dimensional cases. The obtained electric field is later used in calculating the displacement vector by considering {\em the Maxwell stress tensor}. The results of the one-dimensional simulations illustrate that the use of {\em the Maxwell stress tensor} generates a perfect agreement with the analytical model. Since there are no analytical expressions for arbitrary geometries and structures with several materials and space-charge regions, the application of the FE method to such problems results in a better understanding of composite properties that may lead to new tailored materials.  
 
\section{Numerical Modeling}
\label{sec:numerical-modeling}
\begin{figure}[t]
  \centering
  \includegraphics[width=3in]{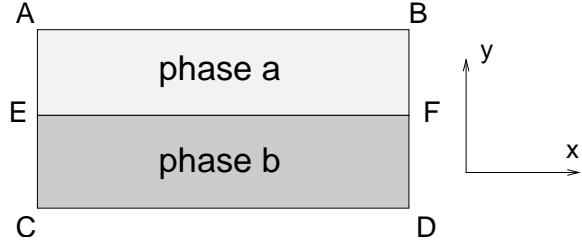}
  \caption{Computation domain for the two-layer system.}
  \label{fig:geom}
\end{figure}
\begin{figure*}[b]
\centering
\begin{tabular*}{6in}{lr}
  \includegraphics[width=3in]{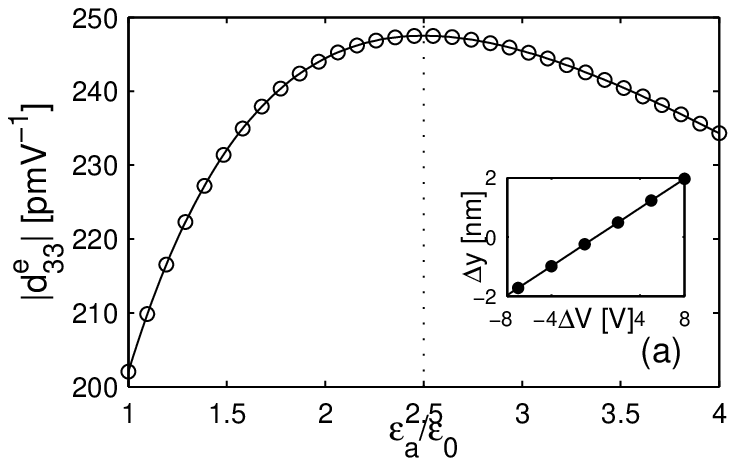}&
  \includegraphics[width=3in]{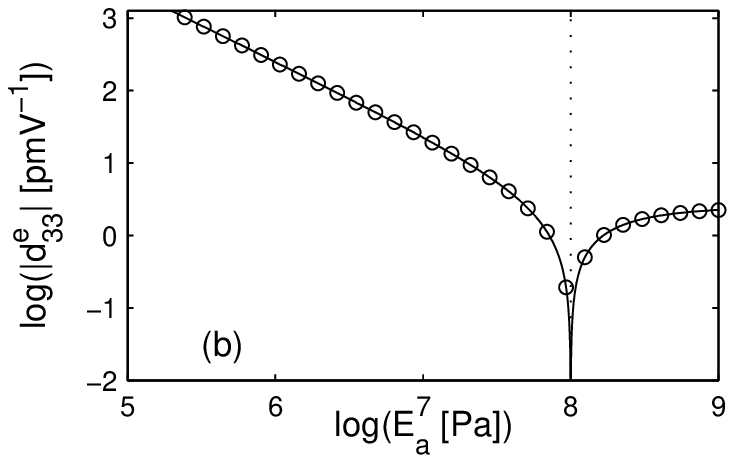}
\end{tabular*}
\caption{Calculated piezoelectric coefficients $d_{33}^{\sf e}$ as a function of (a) the relative permittivity of phase {\sf a} and (b) Young's modulus of phase {\sf a}. The symbols $(\circ)$  and the solid lines (\full) represent the results of the FE calculations and the analytical expression of Eq.~(\ref{eq:1}), respectively. The dotted (\dotted) lines represent the permittivity and Young's modulus of phase {\sf b}, respectively. The inset in (a) illustrates the numerical results for $\varepsilon_{\sf a}=\varepsilon_{\sf_b}$ for various voltages, the slope is the piezoelectric coefficient, $246\ \pico\meter\reciprocal\volt$.}
\label{fig:1}       % Give a unique label
\end{figure*}
A model for the electro-mechanical response of a double-layer dielectric system with an interface charge $\rho$ at the double-layer boundary has been presented by~\citet{Kac0}. Layered systems have been shown as the optimal matrix microstructures for piezocomposites\cite{Torquato1998}. In the model, the dielectric and elastic properties of the phases {\sf a} and {\sf b} with thicknesses $\xa$ and $\xb$ are expressed with the high-frequency relative permittivities $\epsilon_{\sf a,b}$ and Young's moduli $E^{\rm Y}_{\sf a,b}$, respectively. The piezoelectric coefficient for the composite system is then calculated as the ratio of the change of surface charge to the force applied perpendicular to the surface\cite{Kac0};
\begin{equation}
  \label{eq:1}
  d_{33}^{\sf e}=-\frac{\rho\ea \eb \xa \xb (\Ea - \Eb)}{\Ea \Eb (\eb \xa+\ea \xb)^{2}}
\end{equation}
Since there is no intrinsic piezoelectricity assumed in the phases, the coefficient $d_{33}$ is written as an effective material property with a superscript `e'. The term on the right-hand side of Eq.~(\ref{eq:1}) containing permittivities is proportional to the effective dielectric permittivity of the system, $\epsilon^{\sf e}=\ea\eb(\eb\xa+\ea\xb)^{-1}$. For complex structures, or layered systems with materials having different Poisson's ratios, Eq.~(\ref{eq:1}) can not be applied. %The %A parameter $\alpha$ is introduced to show the deviations from the analytical model.

Similarly, applying a numerical model, one should arrive at the same results. To this end, we have employed the FE method and performed simulations  on a double-layer system. %Later a two-dimensional binary mixture is constructed with one of the phases being as a disk. {
In the FE calculations, electrostatic field and  stress-strain relations are solved simultaneously. Neglecting the polarization of the phases, the electric field $\mathbf{E}(=-\nabla{\rm \Phi}({\mathbf r}))$ distribution  is obtained from Poisson's equation:
\begin{equation}
  \label{eq:2}
  -\nabla\cdot\left[ \varepsilon({\mathbf r})\nabla{\rm \Phi}({\mathbf r})\right]=\rho({\mathbf r})
\end{equation}
where ${\rm \Phi}$ is the electric potential distribution, and $\varepsilon$ and $\rho$ are the dielectric permittivity ($\varepsilon=\epsilon\varepsilon_0$, $\varepsilon_0$ is the permittivity of free space and $\varepsilon_0=8.854\ \pico\farad\reciprocal\metre$) and the charge density as space-dependent variables, respectively. For the stress-strain relations, Navier's equation is solved,
\begin{equation}
  \label{eq:3}
    -\nabla \cdot\mathbf{T}({\mathbf r})={\rm K}({\mathbf r}).
\end{equation}
Here $\mathbf{T}$ and ${\rm K}$ are the stress tensor and the body force, respectively. The stress $\mathbf{T}$ is proportional to the gradient of the unknown displacement $\mathbf{u}(\mathbf{r})$ and the space-dependent material coefficient $c$, $\mathbf{T}=c\nabla \mathbf{u}$. The coefficient $c$ is a function of Young's modulus $E^\mathrm{Y}$ and Poisson's ratio $\nu$ of the specific medium. The coupling of the electric field to the mechanical stress is achieved through  {\em the Maxwell stress tensor}\cite{LL,FEMLAB_EMM}. 
\begin{equation}
  \label{eq:4}
  \mathbf{T'}({\mathbf r})=c({\mathbf r})\nabla\mathbf{u}({\mathbf r})-\frac{1}{2}\mathbf{D}({\mathbf r})\cdot\mathbf{E}({\mathbf r})+\mathbf{E}({\mathbf r})\mathbf{D}^{\rm T}({\mathbf r})
\end{equation}
where $\mathbf{T'}$ is the generalized stress tensor. $\mathbf{E}$ and $\mathbf{D}$ are the electric field and the dielectric displacement vectors, and the superscript `T' denotes a transposed matrix. The last term in Eq.~(\ref{eq:4}) is the matrix direct product. Inserting (\ref{eq:4}) into the stress-strain relation of Eq.~(\ref{eq:3}), the general equation for our calculations is obtained:
\begin{equation}
  \label{eq:5}
    -\nabla \cdot[c({\mathbf r}) \nabla \mathbf{u}({\mathbf r})-\frac{1}{2}\mathbf{D}({\mathbf r})\cdot\mathbf{E}({\mathbf r})+\mathbf{E}({\mathbf r})\mathbf{D}^{\rm T}({\mathbf r})]={\rm K}({\mathbf r})
\end{equation}
It is worth mentioning that the spatially varying material parameter $\varepsilon$ and the charge distribution $\rho$ in Eq.~(\ref{eq:2}) are functions of the displacement $\mathbf{u}$\cite{Reimund1983,Lewiner2000}. However,  this coupling is not taken into consideration in this paper.  

For layered structures and for one-dimensional simulations, the  non-diagonal components in the last term on the left-hand side of Eq.~(\ref{eq:5}) are neglected, and the Poisson's ratio $\nu$ of the materials in question is insignificant. No body-force term is considered in the simulations. In the following part, we solve Eqs.~(\ref{eq:2}) and (\ref{eq:5}) with a nonlinear solver based on a commercially available FE software package\cite{FEMLAB}. First two- and three-layer systems with charge distribution(s) perpendicular to the applied field direction (in the $y$-direction) are considered; such a problem is pure one-dimensional. Later the charge distribution is considered to be non-uniform and again perpendicular to the applied field direction, which makes the problem two-dimensional. We have also assumed irregular two-dimensional structures in which the fraction of phase {\sf a} is decreased in the layer.

\section{Results and discussions}
\label{sec:results-discussions}

\subsection{Two-layer system}
\label{sec:two-layered-system}

An application of the numerical model to the geometrical conditions described by \citet{Kac0} confirmed that our approach is valid in one dimension with $\nu_{\sf a,b}\approx0$ and $c=E^{\rm Y}$. In these simulations, the geometry for the computation is a layered structure with phases {\sf a} and {\sf b}, as shown in Fig.~\ref{fig:geom}. The two-layer system is a charged system that produces an external field which leads to piezoelectric properties\cite{Lewiner}.

The mechanical boundary conditions in the simulations are chosen such that the structure is fixed at CD, where it is not allowed to deform in the $x$- and $y$-directions. At the boundaries AB, AC and BD, the structure is allowed to move in the $x$- and $y$-directions. The voltage boundary conditions are applied at CD and AB with voltages $V_{CD}$ and $V_{AB}$, and at AC and BD, symmetry conditions are assumed for the static electrical calculations, $\partial V/\partial x=0$. Moreover, a surface charge density of $1\ \milli\coulombpersquaremetrenp$ (attainable in experiments\cite{Bauer2000,Reimund2000b}) is applied at the internal EF boundary. The thicknesses of the phases are taken equal to each other, $\xa=\xb=1\ \milli\meter$. %It should be pointed out that for thinner layers the thickness change is smaller. 
The layer thickness in the simulations would not affect the overall effective properties of the mixture.


The electric-field-induced mechanical deformation $\Delta y$ can be expressed as a power series. In that case, the electrostrictive $a_{33}^{\sf e}$ and piezoelectric $d_{33}^{\sf e}$ coefficients can be estimated from pairs of the position displacements at the boundary AB and the voltage difference between the AB and CD boundaries, $\Delta V=V_{\rm AB}-V_{\rm CD}$. %The deformation $y$ in the y-direction can be written as
\begin{eqnarray}
  \Delta y&=&y-y_0 \nonumber \\
  &=&\left[ d_{33}^{\sf e} + a_{33}^{\sf e}  \Delta V + \dots  \right] \Delta V \label{eq:6}
\end{eqnarray}
where $y_0$ is the deformation without the voltage applied and due to the charge at the interface EF. We have adapted the first two terms of the series expansion at the right-hand-side of Eq.~(\ref{eq:6}) in the numerical data analysis. The numerical results are shown as an inset in Fig.~\ref{fig:1}a. In the inset, the parameters $y_0$, $d_{33}^{\sf e}$ and $a_{33}^{\sf e}$ are $5.65\times10^{-6}\ [\meter]$, $2.46\times10^{-10}\  [\pico\meter\reciprocal\volt]$ and $-2.13\times10^{-16}  [\pico\meter\rpsquare\volt]$, respectively. Observe that $a_{33}^{\sf e}\cdot \Delta V \ll d_{33}^{\sf e}$. The coefficients can be calculated with the following relations for the nonlinear, $a_{33}^{\sf e}\cdot \Delta V \gg d_{33}^{\sf e}$, and linear $d_{33}^{\sf e}\gg a_{33}^{\sf e}\cdot\Delta V$  regimes, 
\begin{eqnarray}
a_{33}^{\sf e}&=&{\Delta y}{\Delta V^{-2}} \quad [\pico\meter\rpsquare\volt]  \label{eq:7}\\
  d_{33}^{\sf e}&=&{\Delta y}{\Delta V^{-1}} \quad [\pico\meter\reciprocal\volt]  \label{eq:8}
  \end{eqnarray}

In Fig.~\ref{fig:1}, the simulation results and their comparison with Eq.~(\ref{eq:1}) are presented. The permittivity and Young's modulus of phase {\sf b} are kept constant, $\eb=2.5$ and $\Eb=100\ \mega\pascal$, respectively. First, the permittivity of phase {\sf a} is varied between the free-space value $\varepsilon_0$ and $4\varepsilon_0$ while keeping  Young's modulus of phase {\sf a} constant, $\Ea=1\ \mega\pascal$. Then keeping the permittivity of phase {\sf a} constant, $\ea=2.5$, Young's modulus of phase {\sf a} is altered between $100\ \kilo\pascal$ and $1\ \giga\pascal$. These permittivity values are chosen, since most of the polymers have relative permittivities between $2$ and $12$. In addition, traditional materials containing voids (low-$\kappa$ materials) usually have relative permittivities between $1$ and $2$. Young's moduli of the phases are on the other hand assigned on the basis of the effective medium theory\cite{TorquatoBook} for  layered structures and the measured effective moduli of porous polymers from \citet{Bauer2000} and of some solid polymers from {\tt MatWeb}\cite{matweb}.

The results are illustrated in Figs.~\ref{fig:1}a and \ref{fig:1}b.  In Fig.~\ref{fig:1}a, the effective piezoelectric coefficient $|d_{33}^{\sf e}|$ is plotted as a function of the relative permittivity of phase {\sf a} permittivity. The symbols $(\circ)$ and the solid line (\full) represent the results of the FE calculations and the analytical expression Eq.~(\ref{eq:1}), respectively. There is a maximum of $|d_{33}^{\sf e}|$ when the permittivities of the phases are equal to each other, $\ea=\eb$, which is also expected from Eq.~(\ref{eq:1}). This indicates that one should tailor or choose materials with matching dielectric properties. %To achive higher piezoelectric coefficients is strongly dependent on the mechanical properties of the phases and their concentration. 
In Fig.~\ref{fig:1}b the logarithm of $|d_{33}^{\sf e}|$ is displayed as a function of Young's modulus of phase {\sf a} for $\ea=\eb=2.5$. For a stiff phase {\sf a}, $\Ea>\Eb$, $d_{33}^{\sf e}$ converges to a constant value. % that yields a negative $\ddd$. 
As $\Ea$ matches the value of $\Eb$, $\Ea\approx\Eb$ there is no piezoelectric activity, $\log(\Ea)=8$. As the stiffness of phase {\sf a} becomes lower than that of phase {\sf b},  $|d_{33}^{\sf e}|$ increases. For $\Ea\ll\Eb$ there is a linear relation  between $\Ea$ and $|d_{33}^{\sf e}|$ in the log-log graphs. The simulation indicate that there are are no nonlinear effects of the applied voltage on the deformation  in these two-layer simulations, so that $a_{\rm 33}^{\sf e}\approx0$ in Eq.~(\ref{eq:6}), see inset in Fig.~\ref{fig:1}a. All these observation are also expected from Eq.~(\ref{eq:1}), which is also presented in the figures with solid lines (\full).

%\subsection{Influence of charge distribution}
%\label{sec:infl-charge-distr}
\begin{figure}
\centering
  \includegraphics{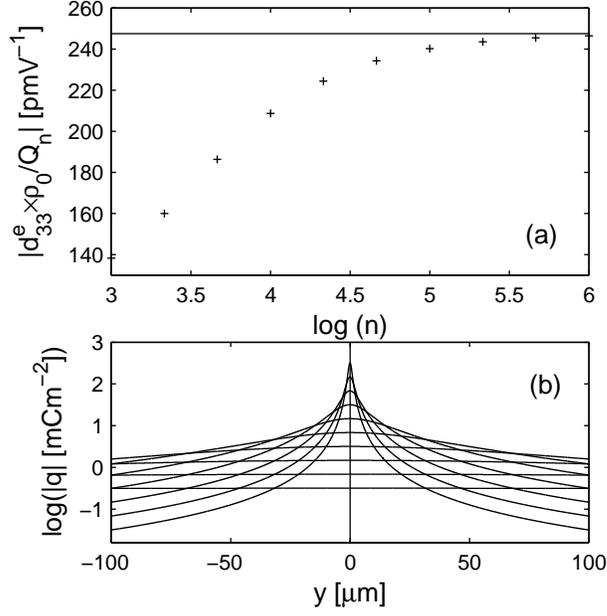}
\caption{(a) Calculated absolute values of piezoelectric coefficient $|d_{33}^{\sf e}|$ normalized to the actual delta surface charge density $\rho_0$ and the total assumed volume charge density $Q_n=\int \rho(y) {\rm d}y$. The symbols $(+)$ are results from the FE calculations, the solid line (\full) represents the actual value from Eq.~(\ref{eq:1}), $|d_{33}^e|=247.5\ \pico\meter\reciprocal\volt$ for a charge sheet. (b) Charge density distributions at the interface for $3\le n\le6$ in Eq.~(\ref{eq:9}. }
\label{fig:delta}       % Give a unique label
\end{figure}

Next, we examine the influence of the charge distribution at the interface on $\ddd$. To this end we introduce a volume charge distribution $\rho(x,y)$ which is only varied in the $y$-direction. The validity of Eq.~(\ref{eq:1}) for such cases is discussed below. The charge distribution is generated with a delta sequence~\cite{Butkov}, 
\begin{equation}
  \label{eq:9}
  \phi_n(y)=n\pi^{-1}[1+n^2 y^2]^{-1}
\end{equation}
where $$\lim_{n\rightarrow\infty}\int \phi_n(y){\rm d}y=1.$$ 
 In Eq.~(\ref{eq:9}), $n$ is a shape parameter. For charge distributions in the form of $$\rho(x,y)=\rho_0\phi_n(y);$$ the volume charge distribution becomes surface charge distribution $\rho=\rho_0\delta(y)$ as $n\rightarrow\infty$. 

In Fig.~\ref{fig:delta}a the calculated $|d_{33}^e|$ is presented as normalized with the volume charge density $\rho_0/Q_n$ for different shape parameters $n$, where 
$$Q_n=\int \rho_0\cdot \phi_n(y) {\rm d}y.$$
The total volume charge density is not equal to $\rho_0$ because of the finite $n$. However, as $n$ gets closer to $10^6$, Eq.~(\ref{eq:9}) can be used to represent a charge sheet. In Fig.~\ref{fig:delta}b volume charge distributions are illustrated for the values of $n$ used to calculate $\ddd$ in Fig.~\ref{fig:delta}a. As the charge penetrates inside the material, forming a true space charge, the piezoelectric response of the composite system becomes weaker compared to a uniform sheet-like surface charge distribution.

\subsection{Three-layer systems}
\label{sec:three-layer-systems}
\begin{table*}
\caption{Minimum and maximum values of the parameters used in Eqs~(\ref{eq:6}) for various simulations. Order means the layer sequence, soft-rigid-soft (SRS) and rigid-soft-rigid (RSR). The parameters varied in the simulations are also presented.}
\label{table1}       % Give a unique label
\centering
%\begin{tabular}{p{1.5in} p{.5in} p{.5in} p{.5in} p{.5in} p{.5in} p{.5in}}
\begin{tabular*}{5in}{l@{\extracolsep{\fill}}rrrrrr}
\hline\noalign{\smallskip} 
Order/ & \multicolumn{2}{c}{$\aaa\ [\femto\meter\rpsquare\volt]$} &
\multicolumn{2}{c}{$\ddd\ [\meter\reciprocal\volt]$} &
\multicolumn{2}{c}{$y_0\ [\micro\meter]$} \\
Parameter & $\min$&$\max$& $\min$&$\max$& $\min$&$\max$\\
\hline\noalign{\smallskip}\hline\noalign{\smallskip} 
\multicolumn{7}{c}{Two-layer system} \\
\hline\noalign{\smallskip} 
\full/$\ea$& & &  $204\pico$ &$250\pico$ & $4.72$ & $5.70$ \\
\full/$\Ea$& & &  $-2.00\pico$ &$2.49\nano$ & $0.07$ & $56.5$ \\
\hline\noalign{\smallskip} 
\multicolumn{7}{c}{bipolar charge system} \\
\hline\noalign{\smallskip} 
RSR/$\ea$ & 2.27 & 2.79 & $-13.9\femto$ & $-11.4\femto$ & 0.23 & 0.25\\
SRS/$\ea$ & 2.27 & 2.80 & $-245\pico$ & $-200\pico$ & 4.66 & 5.61\\
RSR/$\Ea$ & 0.03 & 27.7 & $-2.47\nano$ & $2.20\pico$ & 0.06 & 55.5\\
SRS/$\Ea$ & 0.03 & 27.7 &  $-2.20\pico$ & $2.47\nano$ & 0.06 & 55.5\\
\hline\noalign{\smallskip} 
\multicolumn{7}{c}{mono-polar charge system} \\
\hline\noalign{\smallskip} 
RSR/$\ea$ & 2.27 & 2.79 & $11.3\femto$ & $14.0\femto$ & 13.98 & 55.9\\
SRS/$\ea$ & 2.27 & 2.79 & $-13.9\femto$ & $-11.4\femto$ & 0.23 & 0.25\\
RSR/$\Ea$ & 0.03 & 27.5 & $-0.13\femto$ & $0.15\pico$ & 0.02 & 223.7\\
SRS/$\Ea$ & 0.03 & 27.7 & $ 0.14\pico$ & $0.13\femto$ & 0.02 & 0.32\\
\hline\noalign{\smallskip} 
\end{tabular*}
\end{table*}
\begin{figure}[t]
  \centering
  \includegraphics[width=3in]{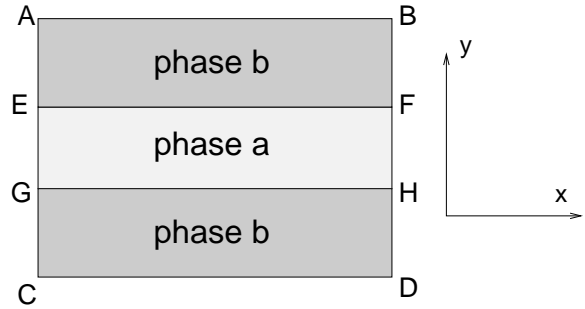}
  \caption{Computation domain for the three-layer system.}
  \label{fig:geomthree}
\end{figure}

\begin{figure}
\centering
  \includegraphics{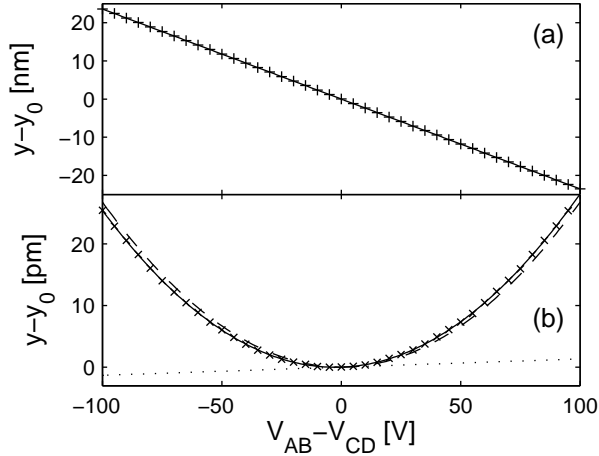}
\caption{Thickness change $y-y_0$ in a three-layer structure for (a) bi- and (b) mono-polar charge distributions as a function of the voltage difference between the electrodes $\Delta V=V_{\rm AB}-V_{\rm CD}$. The symbols are the values calculated with the FE method. The solid lines (\full) are the best linear and quadratic fits. In (b) the fitted curve is separated into its linear (\dotted) and quadratic (\dash) parts, in order to show the electrostrictive and piezoelectric contributions.}
\label{fig:voltage}       % Give a unique label
\end{figure}

In order to enhance the efficiency of an electro-mechanical transducer, it is possible to stack layers of different materials. Such a system can be the three-layer binary structure (`a sandwich')  which allows to employ and investigate different charge polarities. 
%One can further prepare systems with several layers. 
We therefore apply the numerical method to a three-layer binary system as shown in Fig.~\ref{fig:geomthree}. In these simulations, (i) bipolar and (ii) mono-polar charge systems are assumed. In a bipolar system, the charge polarities at interfaces EF and GH in Fig.~\ref{fig:geomthree} are opposite to each other. In a mono-polar system, on the other hand, they have the same polarity. (The main difference between bi- and mono-polar charge systems is that bipolar charge systems do not contain unbalanced internal charge). The deformations $\Delta y$ are analyzed as  functions of the applied voltage difference $V_{\rm AB}-V_{\rm CD}$ with Eq.~(\ref{eq:6}). The deformation $\Delta y$ of a bipolar charge system is dominated by the contribution of the linear term, $\ddd$. The deformation of a mono-polar charge system, on the other hand, shows negligible linear dependence and  the quadratic term, $\aaa$ is important indicating `electrostrictive phenomena'. The calculated deformations for these two systems are shown in Fig.~\ref{fig:voltage}. The deformation with the same applied voltage is 1000 times higher for bipolar  than for mono-polar charge. 

In the simulations, the surface charge densities are $\pm1\ \milli\coulombpersquaremetrenp$  and $1\ \milli\coulombpersquaremetrenp$ for bi- and mono-polar charge distributions. Again, similar to the two-layer system, we first alter the permittivity of the phase {\sf a} keeping Young's moduli of both phases and the permittivity of the phase {\sf b} constant. Later, Young's modulus of phase {\sf a} is varied while all other material parameters are kept constant. The total thicknesses of the phases are equal to each other $\xa=\xb$. There are two different combinations of interest: (i) phase {\sf a} in Fig.~\ref{fig:geomthree} is rigid and the structures is soft-rigid-soft (SRS) or (ii) phase {\sf a} is soft which leads to a rigid-soft-rigid (RSR) structure. The mono-polar-charge cases do not produce significant electro-mechanical activity comparable to the bipolar ones. Experimental observations performed on some prepared RSR structures, have verified the numerical simulations such that the RSR structures do not yield any observable piezoelectric activity. Moreover, mono-polar-charge samples do not show any measurable electro-mecha\-nical activity. 
\begin{figure*}[t]
\centering
\begin{tabular*}{6in}{lr}
  \includegraphics[width=3in]{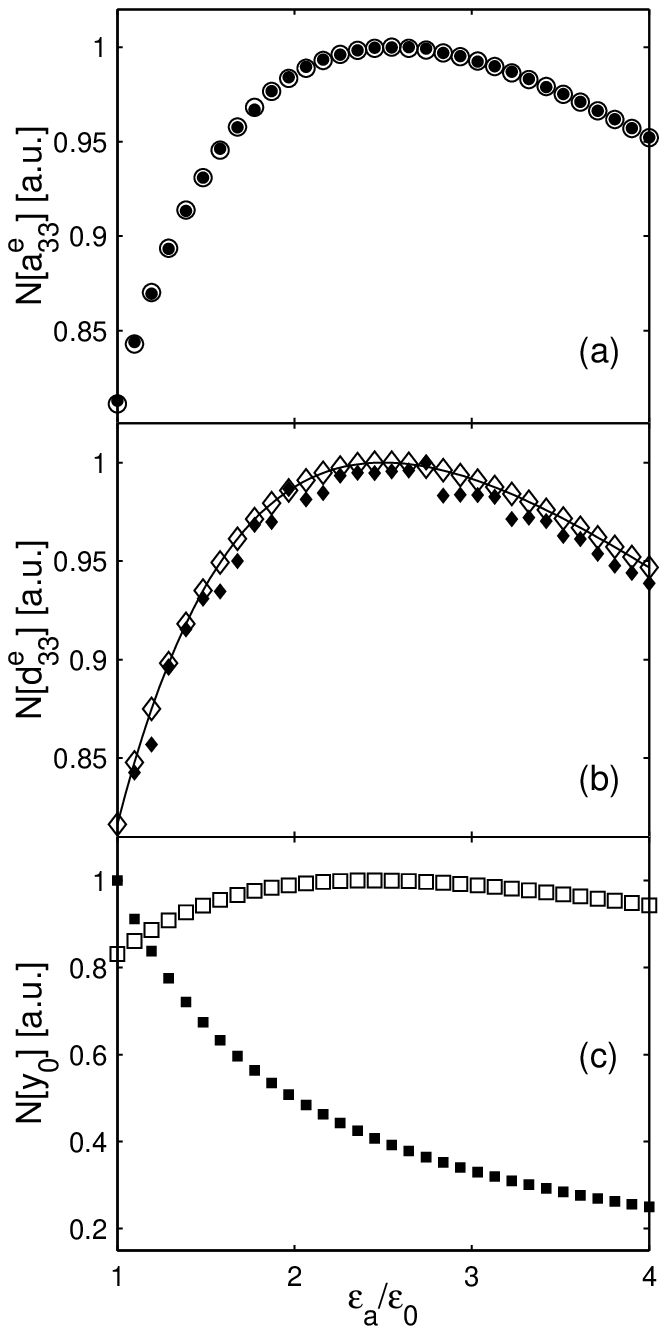}&
  \includegraphics[width=3in]{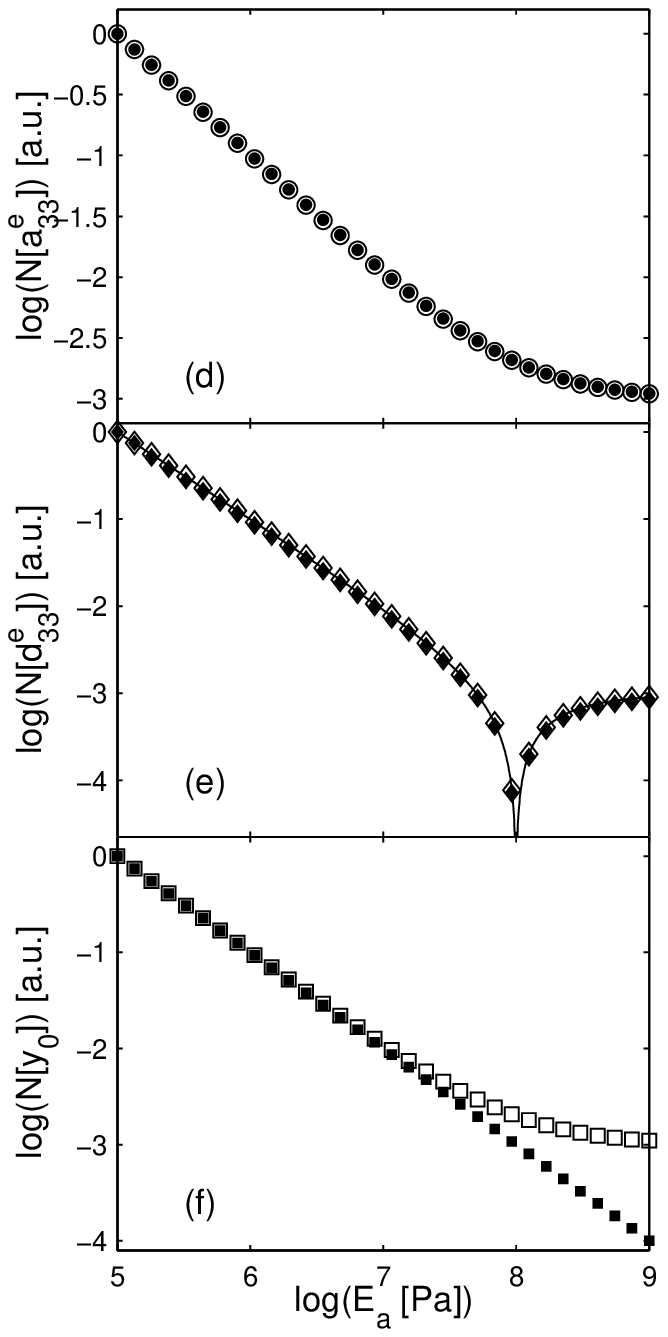}
\end{tabular*}
\caption{Calculated (a) electrostriction $\aaa$ (b) piezoelectric coefficients $\ddd$ and (c) deformation $y_0$ due to interface charge  for altered permittivity of phase {\sf a}. (d) Electrostriction $\aaa$ (e) piezoelectric coefficients $\ddd$ and (f) deformation $y_0$ due to interface charge for altered Young's modulus of phase {\sf a}. The parameters are normalized with respect to their $\max$ or $\min$ values, which are presented in Table~\ref{table1}. The filled and empty symbols represent the mono- and bipolar charge cases for the simulations. The solid lines ($\full$) in (b) and (e) are obtained from Eq.~(\ref{eq:1}) with interface charge  $\rho=970$ and $992\ \micro\coulombpersquaremetrenp$, respectively.}
\label{fig:homohete_eps}       % Give a unique label
\end{figure*}

The results of the simulations are illustrated in Fig.~\ref{fig:homohete_eps}. In the figures, the values are normalized respective to a maxima of the absolute parameter values, $\max|\aaa|$,    $\max|\ddd|$  and  $\max|y_0|$. The normalization values are listed in  Table~\ref{table1} as $\min$ and $\max$. In the table, the rigid and soft phase order of the structures are presented together with the parameter that is varied. While the total thickness change behaves differently, the normalized electrostrictive and piezoelectric coefficients had the same shape as functions of the relative permittivity of phase {\sf a}.

In Figs.~\ref{fig:homohete_eps}b and \ref{fig:homohete_eps}e, solid lines (\full) represent the curve-fitting results obtained with Eq.~(\ref{eq:1}) in which the surface charge $\rho$ is used as a free parameter. It is found that $\pm 1\ \milli\coulombpersquaremetrenp$ on the interfaces of a three-layer system yields the same results as a two-layer system with approx. $1\ \milli\coulombpersquaremetrenp$ at the interface. This is due to the symmetry plane between the two charge layers that can be assumed as in the method of images\cite{Eyges}.  Since there is no external field in a three-layer system with bipolar charge--it is neutral--it is to be preferred for non-electrostatic industrial applications. Except for the results of the RSR/$\ea$ case, the other three bipolar charge systems result in high piezoelectric coefficients, $\ddd>200\ \pico\meter\reciprocal\volt$. It is interesting that the $\max|\ddd|$ and $\min|\ddd|$ values obtained for the RSR/$\ea$ and SRS/$\ea$ cases are not comparable. The RSR structure is not suitable for any electro-mechanical application. The reason for the very low piezoelectric coefficients in the RSR case is that the structure expands or contracts in response to the internal charge distribution, even in the absence of an applied voltage--$y_0$ in Table~\ref{table1}. After this deformation there is no (or very little) space for the structure to respond to the applied voltage.  It is observed that a $\ddd$ value on the order of $\nano\meter\reciprocal\volt$ can be obtained if materials with low Young's modulus ($E^{\rm Y}<1\ \mega\pascal$) are employed such as non-conducting polymer foams and natural cork\cite{Ashby}. %, however the desired foam should have elastic properties. 

%\begin{figure}
%\centering
%\caption{Calculated (a) electrostriction $\aaa$ (b) piezoelectric coefficients $\ddd$ and (c) deformation $y_0$ due to interface charge  for altered permittivity of phase {\sf a}. The parameters are normalized with respect to their $\max$ or $\min$ values. The filled and empty symbols represent the homo- and bipolar charge cases for the simulations. The solid line ($\full$) in (b) is the curve-fit to bipolar charge ($\diamond$) case using simulation material parameters with $992\ \micro\coulombpersquaremetrenp$ charge at the interface.}
%\label{fig:homohete_you}       % Give a unique label
%\end{figure}
\begin{figure}[t]
  \centering
  \includegraphics{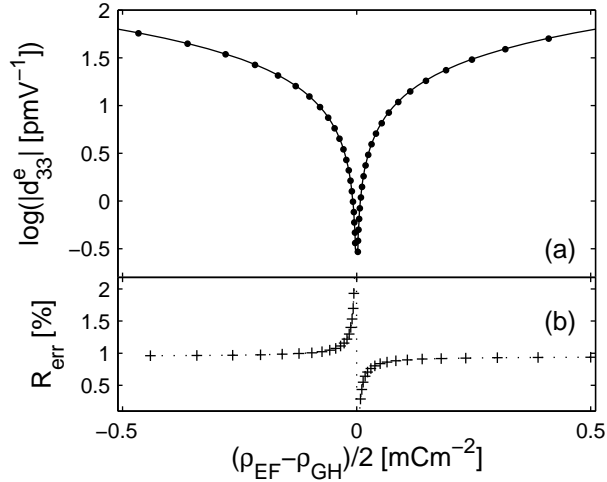}
  \caption{(a) Calculated piezoelectric coefficient $|\ddd |$ ($\bullet$) as a function of difference in total charges at boundaries EF and GH. The solid line (\full) is Eq.~(\ref{eq:1}) with half the apparent charge. (b) Relative error in percent.}
  \label{fig:homo_vs_charge}
\end{figure}

As mentioned previously, the mono-polar charge systems exhibit very little electro-mechanical activity when the charges at the surfaces are equal. However, at high applied voltages, the electrostrictive effect is significant. To have the same magnitude of the piezoelectric coefficient in a two-layer system one should introduce a  surface charge of about $1\nano \coulombpersquaremetrenp$ at the interface, which is extremely low. If the surface charges are not equal, but have the same sign, the response of the structure can be tailored to be linear in the applied field for specially selected surface charge values. This is indicated in Fig.~\ref{fig:homo_vs_charge}. When the unbalanced charge of the structure is close to zero, $(\rho_{\rm EF}- \rho_{\rm GH})/2\approx0$, the analytical solution and the numerical results deviate,  due to the numerical error. This is an interesting case, since in phases with significant conductivity, charges on the surfaces would recombine. This process would directly affect the electro-mechanical activity and lower $\ddd$.  

\subsection{Two-dimensional system}
\subsubsection{Non-uniform charge density at the interface}
\begin{figure}[t]
  \centering
\begin{tabular*}{6in}{lr}
  \includegraphics{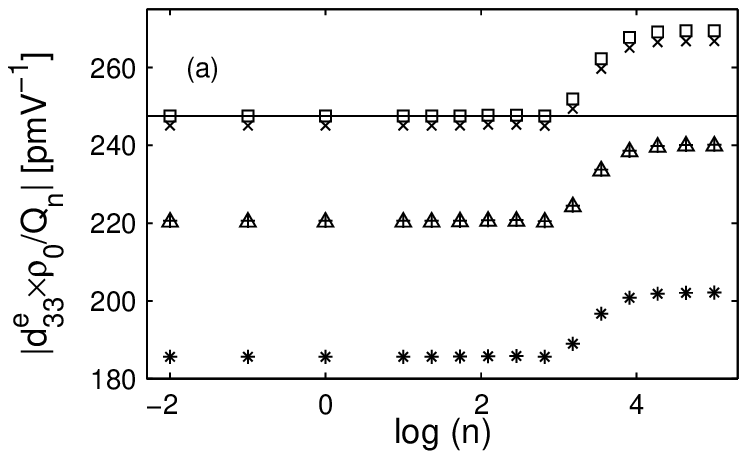}&  
  \includegraphics{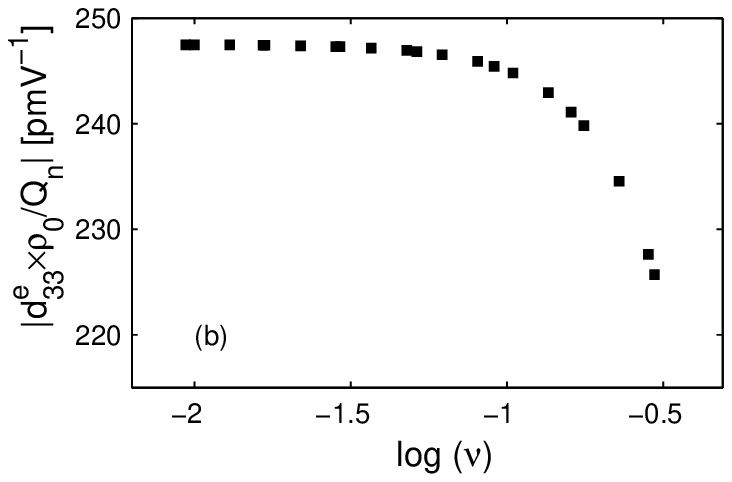}
\end{tabular*}
  \caption{(a) Calculated absolute values of piezoelectric coefficient $|\ddd|$ normalized to the actual two dimensional charge density $\rho(x)$ for various Poisson's ratio of phases ($\nu=\nua=\nub$); $(\Box)$ $\nu\ll0.1$, $(\times)$ $\nu=0.1$, $(+)$ $\nu=0.33$, $(*)$ $\nu=0.5$ and $(\triangle)$ $\nua\ll0.33$ $\nub=0.1$. The solid line (\full) is the value calculated in \S~\ref{sec:two-layered-system} for uniform surface charge distribution. (b) Influence of Poisson's ratio on $\ddd$ for $\nua=\nub\ll0.1$.}
  \label{fig:2D}
\end{figure}

As an illustration for a two-dimensional solution, we again consider a two-layer system as in Fig.~\ref{fig:geom}. However, this time we just alter the charge distribution at the interface EF in the $x$-direction.
\begin{equation}
  \label{eq:10}
  \rho(x)=\rho_0n\pi^{-1}[1+n^2 x^2]^{-1}
\end{equation}
where $n$ defines the shape of the distribution---for $n\rightarrow0$ surface and $n\rightarrow\infty$ line charge distributions are obtained---, and $\rho_0$ is the charge amplitude. The material properties are taken as $\ea=\eb=2.5$ and $\Ea=1\ \mega\pascal$ and $\Eb=100\ \mega\pascal$. The boundary conditions are assigned as in the previous problem (see \S~\ref{sec:two-layered-system}). Since the problem is now in two-dimensions the Poisson's ratio of phases $\nua$ and $\nub$ influence the resulting effective piezoelectric coefficient $\ddd$. In Fig.~\ref{fig:2D}a, the results obtained for $\ddd$ are illustrated as a function of $n$ for various Poisson's ratios. For small values of n ($n<10^3$) and $\nua=\nub\ll0.1$, $\ddd$ is the same as the one-dimensional simulations, a uniform charge distribution. However, as the charge distribution is altered by changing $n$, the resulting $\ddd$ increases for $n>10^3$ and approaches a constant for $n>10^4$. These higher values of $\ddd$ are expected due to  the electric field distribution in the system, which is enhanced at the tip of the charge distribution (sheet-like discontinuous surface charge in the $x$-direction). The enhancement is higher as the charge distribution approaches a line charge distribution, the piezoelectric coefficient being approx. 10\% higher than that of uniform sheet charge distribution. %In addition, the influence of the Poisson's ratio is reciprocal on the $\ddd$. This is illustrated i
In Fig.~\ref{fig:2D}a and ~\ref{fig:2D}b show that $\ddd$ decreases with increasing Poisson's ratio, when the latter approaches 0.5. %, $\ddd$ decreases because of the shear strain.

\begin{figure}[t]
  \centering
   \psfragscanon
  \psfrag{4m}[lb]{\sf 0.004}
 \includegraphics[width=4in]{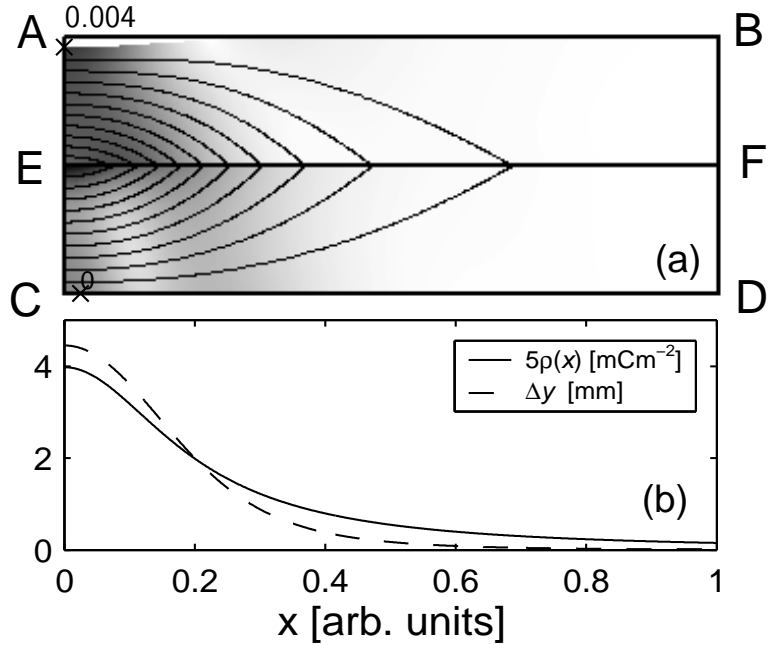}
  \psfragscanoff
  \caption{(a) Mechanical stress (von Mises stress) in gray-scale  and voltage distribution in contour plot (for $\Delta V=0\ \volt$). The structure is deformed, the maximum and minimum deformations are marked with symbols ($\times$), and the values are normalized to thickness of the whole structure, AC. (b) Charge density at the interface EF and deformation of the boundary AB. The charge density is calculated for $n=10$ and $\rho_0=1\ \milli\coulomb\rpsquare\meter$ in Eq.~(\ref{eq:10}).}
  \label{fig:2Dpic}
\end{figure}

Finally, as an illustrative example in Fig.~\ref{fig:2Dpic}a the stress distribution and electrical potential are shown. The voltage difference between the AB and CD boundaries is $0\ \volt$. The Poisson's ratio of the phases are as follows, $\nua=0.33$ $\nub\ll0.1$. In Fig.~\ref{fig:2Dpic}b the charge distribution at the interface and the deformation on the boundary AB are presented. The deformation is purely due to the interfacial charge, and it is larger than the previous cases considered. The deformation is localized as the charge distribution.
%\begin{table}
%\caption{Material parameters in the simulations.}
%\begin{tabular}{lccc}
%\hline\noalign{\smallskip} 
%      & permittivity & Young's modulus & Poisson's ratio  \\
%Phase & $(\times\varepsilon_0)$ & $\log(E)$ & $\nu$  \\
%{\sf a} & 1-4 &  5-9 & $\sim0$\\
%{\sf b} & 2.5 &  8 & $\sim0$\\
%\end{tabular}
%% Or use
%\vspace*{5cm}  % with the correct table height
%\end{table}
%Since the Eq.~(\ref{eq:1}) does not take into account the Poisson's ratio. it is neglected, $\nu\approx0$, that yield $D=E$. 

\subsubsection{Irregularly shape sample}\label{sec:irreg-shape-sample}
\begin{figure}[t]
  \centering
  \includegraphics{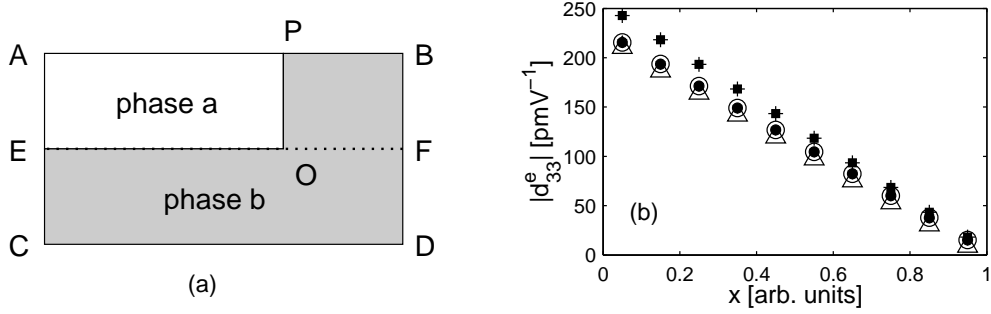}
  \caption{(a) The geometry used in the calculations for the irregular shape. Phase {\sf b} is considered to be soft, the length PB is used as a parameter in the calculations, $x=|{\rm PB}|/|{\rm CD}|$. The dotted line between EF show the charge interface. (b) Calculated absolute values of piezoelectric coefficient $|\ddd|$ for various PB ratio and Poisson' ratio when the charge distribution of $1\ \milli\coulomb\rpsquare\meter$ is considered in the EF interface; $(\blacksquare)$ $\nua=0.33$ and $\nub\ll0.1$, $(+)$ $\nua\ll0.1$ and $\nub\ll0.1$, $(\bullet)$ $\nua\ll0.1$ and $\nub=0.33$, $(\triangledown)$ $\nua=0.33$ and $\nub=0.33$, $(\circ)$ $\nua=0.5$ and $\nub=0.33$. When only charge is considered in the EO the piezoelectric  coefficient is not altered significantly, $(\triangle)$ $\nua=0.5$ and $\nub=0.33$. }
  \label{fig:ref2}
\end{figure}

\begin{figure}[t]
  \centering
  \psfragscanon
  \psfrag{2m}[cb]{\sf 0.002}
  \includegraphics[width=5in]{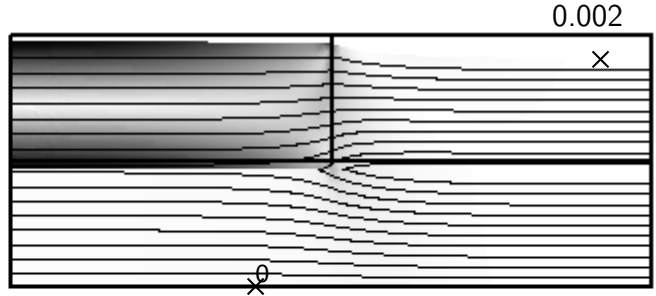}
  \psfragscanoff
  \caption{Mechanical stress (von Mises stress) in gray-scale  and voltage distribution in contour plot (for $\Delta V=0\ \volt$). The structure is deformed, the maximum and minimum deformations are marked with symbols ($\times$), and the values are normalized to the thickness of the whole structure, AC. The Poisson's ratio of phase {\sf a} is 0.33.}
  \label{fig:2dref2}
\end{figure}

In this section we have considered an irregular shape as illustrated in Fig.~\ref{fig:ref2}a. Phase {\sf a} is considered to be changing its width, $|{\rm  AP}|$. In the simulation, the same consideration as in the previous cases are considered, $\ea=\eb=2.5$ and $\Ea=1\ \mega\pascal$ and $\Eb=100\ \mega\pascal$. The charge is assumed to be on the interface EF with the charge density $1\ \milli\coulomb\rpsquare\meter$. The results are plotted in Fig.~\ref{fig:ref2}b for various Poisson's ratio of phases and fraction of phase {\sf a}, $1-x$ where $x$ is the length of PB.  It is clear that the Poisson's ratio of the soft phase (phase {\sf b}) influences the piezoelectric coefficient $|\ddd|$. Moreover, as the fraction of the rigid phase is lowered, $x\rightarrow0$, the piezoelectric activity is decreasing linearly with the fraction of phase {\sf b}, which is expected. In addition the decrease in $\ddd$ with respect to $x$ is proportional to the Poisson's ratio of phase {\sf b}. In Fig.~\ref{fig:2dref2} the stress and the electric potential distributions are presented as gray-scale and contour plots. %It is illustrative to see the mechanical stress in phase {\sf a} due to the surface charge distributions.

\section{Conclusions}
\label{sec:conclusion}
Numerical simulations on the electro-mechanical properties of layered structures containing charges are reported. The simulations  take the coupling of the electrical and mechanical stresses into account by means of {\em the Maxwell stress tensor}. The results are compared with an analytical model, and it is observed that there is  good agreement for two-layer systems. %When the arrangement for the bipolar-charge three-layer systems are adjusted as a two-layer systems, the analytical model for two-layer system is able to express their electro-mechanical properties. 
The piezoelectricity in three-layer systems shows a strong dependence on the polarity of interface charges. It is concluded that bipolar charge systems clearly yield higher piezoelectric coefficient values than mono-polar charge systems. The mono-polar charge  cases are dominated by `electrostrictive effects'. 

In this paper, we have presented a way of calculating electro-mechanical activity of layered electrets with uniform and non-uniform charge distributions. The numerical model is also applied to a two-layer system with discontinuous interface charge layer. In such a simulation unlike the one-dimensional calculations the Poisson's ratio of the phases become significant. High Poisson's ratio lead to low the electro-mechanical coefficient. As the sheet of charge at the interface is modified toward a line charge distribution the piezoelectric coefficient is increased approx. 10\%. The two-dimensional structures considered have also illustrated the stress distribution in the constituents are of importance, the stiff phase is mechanically stressed due to electromechanical activity in the soft phase. We might conclude that the presence of excess charges (space charge) in heterogeneous systems is crucial for aging of the materials.
%The presented numerical model can be applied to problems with more complex geometries, even effects like a density-dependent permittivity, Young's modulus, Poisson ratio, a significant conductivity or frequency dependent properties can be included. 

As a concluding remark, it is illustrated that in heterogeneous materials, fluctuations in the local electric field could lead to mechanical deformations even if there is no intrinsic piezoelectricity in the phases. The considered cases show that the deformation follows the actual charge distribution at the interface, and distribution of charges plays an important role in materials with irregularity.

\bibliographystyle{unsrtnat}
\bibliography{pie_els}

\begin{thebibliography}{28}
\providecommand{\natexlab}[1]{#1}
\providecommand{\url}[1]{\texttt{#1}}
\expandafter\ifx\csname urlstyle\endcsname\relax
  \providecommand{\doi}[1]{doi: #1}\else
  \providecommand{\doi}{doi: \begingroup \urlstyle{rm}\Url}\fi

\bibitem[Gerhard-Multhaupt(2002)]{ReimundRev}
R.~Gerhard-Multhaupt.
\newblock Less can be more. holes in polymers lead to a new paradigm of
  piezoelectric materials for electret transducers.
\newblock \emph{IEEE Transactions on Dielectrics and Electrical Insulation},
  9\penalty0 (5):\penalty0 850--859, 2002.

\bibitem[Bauer et~al.(2004)Bauer, Gerhard-Multhaupt, and
  Sessler]{ReimundPhysTdy}
S.~Bauer, R.~Gerhard-Multhaupt, and G.~M. Sessler.
\newblock Ferroelectrets: Soft electro active foams for transducers.
\newblock \emph{Physics Today}, pages 37--43, February 2004.

\bibitem[Cady(1964)]{Cady12}
W.~G. Cady.
\newblock \emph{Piezoelectricity: An introduction to the theory and
  applications of electrical phenomena in crystals}.
\newblock Dover Publications Inc., New York, new revised edition, 1964.

\bibitem[Dreyfus and Lewiner(1976)]{Lewiner}
G.~Dreyfus and J.~Lewiner.
\newblock Free energy of electrets.
\newblock \emph{Physical Review B}, 14\penalty0 (12):\penalty0 5451--5457,
  1976.

\bibitem[Wintle and D{\"o}rsam(1989)]{Wintle1989}
H.~J. Wintle and R.~D{\"o}rsam.
\newblock Phenomenological piezoelectricity of polymer insulators.
\newblock \emph{Physical Review B}, 39\penalty0 (6):\penalty0 3862--3870, 1989.

\bibitem[Kacprzyk et~al.(1995)Kacprzyk, Motyl, Gajewski, and Pasternak]{Kac0}
R.~Kacprzyk, E.~Motyl, J.~B. Gajewski, and A.~Pasternak.
\newblock Piezoelectric properties of nonuniform electrets.
\newblock \emph{Journal of Electrostatics}, 35:\penalty0 161--166, 1995.

\bibitem[Paajanen et~al.(2000{\natexlab{a}})Paajanen, V{\"a}lim{\"a}ki, and
  Lekkala]{Lekkala2000a}
M.~Paajanen, H.~V{\"a}lim{\"a}ki, and J.~Lekkala.
\newblock Modelling the electromechanical film ({EMF}i).
\newblock \emph{Journal of Electrostatics}, 48:\penalty0 193--204,
  2000{\natexlab{a}}.

\bibitem[Gerhard-Multhaupt et~al.(2000)Gerhard-Multhaupt, K{\"u}nstler,
  G{\"o}rne, Pucher, Weinhold, Sei{\ss}, Xia, Wedel, and Danz]{Reimund2000b}
R.~Gerhard-Multhaupt, W.~K{\"u}nstler, T.~G{\"o}rne, A.~Pucher, T.~Weinhold,
  M.~Sei{\ss}, Zhongfu Xia, A.~Wedel, and R.~Danz.
\newblock Porous {PTFE} space-charge electrets for piezoelectric applications.
\newblock \emph{IEEE Transactions on Dielectrics and Electrical Insulation},
  7\penalty0 (4):\penalty0 480--492, 2000.

\bibitem[K{\"u}nstler et~al.(2000)K{\"u}nstler, Xia, Weinhold, Pucher, and
  Gerhard-Multhaupt]{Reimund2000a}
W.~K{\"u}nstler, Z.~Xia, T.~Weinhold, A.~Pucher, and R.~Gerhard-Multhaupt.
\newblock Piezoelectricity of porous polytetrafluoroethylene single- and
  multi-film electrets containing high charge densities of both polarities.
\newblock \emph{Applied Physics A}, 70:\penalty0 5--8, 2000.

\bibitem[Sessler and Hillenbrand(1999)]{Sessler1999}
G.~M. Sessler and J.~Hillenbrand.
\newblock Electromechanical response of cellular electret films.
\newblock \emph{Applied Physics Letters}, 75\penalty0 (21):\penalty0
  3405--3407, 1999.

\bibitem[Lindner et~al.(2002)Lindner, Bauer-Gogonea, Bauer, Paajanen, and
  Raukola]{Lindner}
M.~Lindner, S.~Bauer-Gogonea, S.~Bauer, M.~Paajanen, and J.~Raukola.
\newblock Dielectric barrier microdischarges: Mechanisims for charging of
  cellular piezoelectric polymers.
\newblock \emph{Journal of Applied Physics}, 91\penalty0 (8):\penalty0
  5283--5287, 2002.

\bibitem[Schw{\"o}diauer et~al.(2000)Schw{\"o}diauer, Neugschwandtner,
  Schrattbauer, Lindner, Vieytes, Bauer-Gogonea, and Bauer]{Bauer2000b}
R.~Schw{\"o}diauer, G.~S. Neugschwandtner, K.~Schrattbauer, M.~Lindner,
  M.~Vieytes, S.~Bauer-Gogonea, and S.~Bauer.
\newblock Prepareation and characterization of noval piezoelectric and
  pyroelectric polymer electrets.
\newblock \emph{IEEE Transactions on Dielectrics and Electrical Insulation},
  7\penalty0 (4):\penalty0 578--586, 2000.

\bibitem[Paajanen et~al.(2000{\natexlab{b}})Paajanen, Lekkala, and
  Kirjavainen]{Lekkala2000}
M.~Paajanen, J.~Lekkala, and K.~Kirjavainen.
\newblock Electromechanical film {EMF}i-a new multipurpose electret material.
\newblock \emph{Sensors and Actuators A: Physical}, 84:\penalty0 95--102,
  2000{\natexlab{b}}.

\bibitem[Paajanen et~al.(2001)Paajanen, Lekkala, and
  V{\"a}lim{\"a}ki]{Lekkala2001}
M.~Paajanen, J.~Lekkala, and H.~V{\"a}lim{\"a}ki.
\newblock Electromechanical modeling and properties of the electret film
  {EMF}i.
\newblock \emph{IEEE Transactions on Dielectrics and Electrical Insulation},
  8\penalty0 (4):\penalty0 629--636, 2001.

\bibitem[Torquato(2001)]{TorquatoBook}
S.~Torquato.
\newblock \emph{Random Heterogeneous Materials: Microstructure and macroscopic
  properties}, volume~16.
\newblock Springer-Verlag, Berlin, 2001.

\bibitem[Tuncer et~al.(2002{\natexlab{a}})Tuncer, Guba{\'n}ski, and
  Nettelblad]{Tuncer2002b}
E.~Tuncer, S.~M. Guba{\'n}ski, and B.~Nettelblad.
\newblock Non-debye dielectric relaxation in binary dielectric mixtures
  (50-50): {R}andomness and regularity in mixture topology.
\newblock \emph{Journal of Applied Physics}, 92\penalty0 (8):\penalty0
  4612--4624, 2002{\natexlab{a}}.

\bibitem[Tuncer et~al.(2002{\natexlab{b}})Tuncer, Serdyuk, and
  Gubanski]{Tuncer2002a}
E.~Tuncer, Y.~V. Serdyuk, and S.~M. Gubanski.
\newblock Dielectric mixtures: electrical properties and modeling.
\newblock \emph{IEEE Transactions on Dielectrics and Electrical Insulation},
  9\penalty0 (5):\penalty0 809--828, 2002{\natexlab{b}}.

\bibitem[Sigmund et~al.(1998)Sigmund, Torquato, and Aksay]{Torquato1998}
O.~Sigmund, S.~Torquato, and I.~A. Aksay.
\newblock On the design of 1-3 piezocomposites using topology optimization.
\newblock \emph{Journal of Materials Research}, 13\penalty0 (4):\penalty0
  1038--1048, 1998.

\bibitem[Landau and Lifshitz(1982)]{LL}
L.D. Landau and E.M. Lifshitz.
\newblock \emph{Electrodynamics of continuous media}, volume~8 of \emph{Course
  of Theoretical Physics}.
\newblock Perganom Press, New York, 2nd edition, 1982.

\bibitem[Com(2002)]{FEMLAB_EMM}
\emph{{\sc Femlab} Electromagnetics Module}.
\newblock Comsol AB, Stockholm, Sweden, 2.3 edition, 2002.

\bibitem[Gerhard-Multhaupt(1983)]{Reimund1983}
R.~Gerhard-Multhaupt.
\newblock Analysis of pressure-wave methods for the nondestructive
  determination of spatial charge or field distribution in dielectrics.
\newblock \emph{Physical Review B}, 27\penalty0 (4):\penalty0 2494--2503, 1983.

\bibitem[Paris et~al.(2000)Paris, Lewiner, Ditchi, Hol{\'e}, and
  Alqui{\'e}]{Lewiner2000}
O.~Paris, J.~Lewiner, T.~Ditchi, S.~Hol{\'e}, and C.~Alqui{\'e}.
\newblock A finite element method for the determination of space charge
  distributions in complex geometry.
\newblock \emph{IEEE Transactions on Dielectrics and Electrical Insulation},
  7\penalty0 (4):\penalty0 556--560, 2000.

\bibitem[Com(2001)]{FEMLAB}
\emph{{\sc Femlab} User's Guide and Introduction}.
\newblock Comsol AB, Stockholm, Sweden, 2.2 edition, 2001.

\bibitem[Neugschwandtner et~al.(2000)Neugschwandtner, Schw{\"o}diauer,
  Bauer-Gogonea, and Bauer]{Bauer2000}
G.~S. Neugschwandtner, R.~Schw{\"o}diauer, S.~Bauer-Gogonea, and S.~Bauer.
\newblock Large piezoelectric effects in charged, heterogeneous fluoropolymer
  electrets.
\newblock \emph{Applied Physics A}, 70:\penalty0 1--4, 2000.

\bibitem[mat(2003)]{matweb}
Matweb: Material property data.
\newblock Automation Creations, Inc., Christiansburg, VA USA: {\tt
  http://www.matweb.com}, 2003.

\bibitem[Butkov(1968)]{Butkov}
E.~Butkov.
\newblock \emph{Mathematical Physics}.
\newblock Addison-Wesley Series in Advanced Physics. Addison-Wesley Publishing
  Company, Menlo Park, 1968.

\bibitem[Eyges(1972)]{Eyges}
L.~Eyges.
\newblock \emph{The Classical Electromagnetic Field}.
\newblock Dover, New York, 1972.

\bibitem[Ashby and Johnson(2002)]{Ashby}
M.~F. Ashby and Kara Johnson.
\newblock \emph{Materials and Design: The Art and Science of Materials
  Selection in Product Design}.
\newblock Butterworth Heinemann, Oxford, 2002.

\end{thebibliography}
%\begin{thebibliography}{}
%%
%% and use \bibitem to create references.
%Author, Journal \textbf{Volume,} (year) page numbers.
%Author, \textit{Book title} (Publisher, place year) page numbers
%\end{thebibliography}

\end{document}